\documentclass[sigconf]{acmart} 

\copyrightyear{2026}
\acmConference[WWW '26] {Proceedings of the ACM Web Conference 2026}{April 13--17, 2026}{Dubai, United Arab Emirates.}
\acmBooktitle{Proceedings of the ACM Web Conference 2026 (WWW '26), April 13--17, 2026, Dubai, United Arab Emirates}

\usepackage{amsmath,amssymb,amsfonts}
\usepackage{algorithm}
\usepackage{algpseudocode}
\usepackage{algpseudocodex}
\usepackage{enumitem}
\usepackage{graphicx}
\usepackage{textcomp}
\usepackage{xcolor}
\usepackage{amsthm}
\usepackage{amsmath}
\usepackage{url}
\usepackage{soul}
\usepackage{balance}
\usepackage{enumitem}
\usepackage{dsfont}
\usepackage{subcaption}
\usepackage{multirow} 
\usepackage{adjustbox}
\usepackage{array} 

\usepackage{bbm}

\def\BibTeX{{\rm B\kern-.05em{\sc i\kern-.025em b}\kern-.08em
    T\kern-.1667em\lower.7ex\hbox{E}\kern-.125emX}}

\newcommand*{\circled}[1]{\lower.7ex\hbox{\tikz\draw (0pt, 0pt)%
    circle (.5em) node {\makebox[1em][c]{\small #1}};}}
\robustify{\circled}

\begin{document}
\title{Rethink Web Service Resilience in Space: A Radiation-Aware and Sustainable Transmission Solution}
\settopmatter{authorsperrow=4}
 \author{Long Chen}\thanks{Long Chen and Hao Fang contributed equally to this work.} 
 \thanks{Corresponding author: Xiaoyi Fan.}
\affiliation{
\institution{Simon Fraser University}
\city{Burnaby}
\country{Canada}
}
\email{longchen.cs@ieee.org}

\author{Hao Fang}
\affiliation{
\institution{Simon Fraser University}
\city{Burnaby}
\country{Canada}
}
\email{fanghaof@sfu.ca}

\author{Yi Ching Chou}
\affiliation{
\institution{Simon Fraser University}
\city{Burnaby}
\country{Canada}
}
\email{ycchou@sfu.ca}

\author{Haoyuan Zhao}
\affiliation{
\institution{Simon Fraser University}
\city{Burnaby}
\country{Canada}
}
\email{hza127@sfu.ca}

\author{Xiaoyi Fan}
\affiliation{
\institution{Jiangxing Intelligence Inc.}
\city{Shenzhen}
\country{China}
}
\email{xiaoyi.fan@ieee.org}

\author{Zhe Chen}
\affiliation{
\institution{Fudan University}
\city{Shanghai}
\country{China}
}
\email{zhechen@fudan.edu.cn}

\author{Hengzhi Wang}
\affiliation{
\institution{Shenzhen University}
\city{Shenzhen}
\country{China}
}
\email{whz@szu.edu.cn}

\author{Jiangchuan Liu}
\affiliation{
\institution{Simon Fraser University}
\city{Burnaby}
\country{Canada}
}
\email{jcliu@sfu.ca}

\renewcommand{\shortauthors}{Long Chen et al.}

\ccsdesc[500]{Networks~Network services}
\ccsdesc[500]{Networks~Network management}

\keywords{web service resilience; radiation-aware transmission; low-Earth-orbit satellite networks}

\begin{abstract}
Low Earth Orbit (LEO) satellite networks such as Starlink and Project Kuiper are increasingly integrated with cloud infrastructures, forming an important internet backbone for global web services. By extending connectivity to remote regions, oceans, and disaster zones, these networks enable reliable access to applications ranging from real-time WebRTC communication to emergency response portals. Yet the resilience of these web services is threatened by space radiation: it degrades hardware, drains batteries, and disrupts continuity, even if the space-cloud integrated providers use machine learning to analyze space weather and radiation data. Specifically, conventional fixes like altitude adjustments and thermal annealing consume energy; neglecting this energy use results in deep discharge and faster battery aging, whereas sleep modes risk abrupt web session interruptions. Efficient network-layer mitigation remains a critical gap. We propose RALT (Radiation-Aware LEO Transmission), a control-plane solution that dynamically reroutes traffic during radiation events, accounting for energy constraints to minimize battery degradation and sustain service performance. Our work shows that unlocking space-based web services’ full potential for global reliable connectivity requires rethinking resilience through the lens of the space environment itself.
\end{abstract}

\maketitle


\section{Introduction}
\label{sec:introduction}
Low Earth Orbit (LEO) satellite networks have become the backbone of ubiquitous web service access, extending coverage to remote regions, oceans, and disaster zones where ground infrastructure is absent \cite{zhang2025spache}. Leading cloud providers such as Amazon Web Services (AWS) have vertically integrated LEO into their operational backbone through Project Kuiper, which combines a 3,000+ satellite constellation with AWS-powered ground infrastructures\footnote{\url{https://aws.amazon.com/aerospace-and-satellite/}}.
These networks promise to deliver consistent daily web experiences, including webpage browsing, video streaming, real-time WebRTC communications, and emergency rescue portals, and they align with the core demand for accessible and reliable web services globally.

Yet this promise faces a critical gap: existing research on LEO-based web services mostly focuses on mitigating satellite dynamicity like topology changes to tweak latency or throughput \cite{zhang2025spache, zeqilai_inorbit_2024,zhao2023first}. It overlooks a more fundamental service disruption driver, the space environment itself, specifically space radiation, which exerts inherent, long-term impacts on service resilience that cannot be addressed by optimizing satellite motion alone. This radiation-driven threat endangers the resilience infrastructure that cloud providers rely on to sustain daily web access. Existing research’s oversight of this key issue demands a rethink: to enable robust web services in LEO networks, we must shift focus from satellite dynamicity to the space environment’s root threats, starting with radiation.

Radiation impacts on satellites directly harm web services through three key mechanisms \cite{NASASolarEffect}: \emph{atmospheric drag} intensifies with radiation, forcing satellites to consume more battery power for orbit maintenance which prolongs web request latency (e.g., page load delays) and risks service outages \cite{yu2022simulation}; \emph{Total Ionizing Dose (TID) and Total Non-Ionizing Dose (TNID) effects} cause cumulative electronic degradation, increasing data transmission errors that trigger retransmissions of web content (e.g., video streams, large files) \cite{srour2013displacement}; \emph{Single Event Effects (SEE)} from high-energy particles induce sudden electronic disruptions, forcing satellites into sleep mode and interrupting real-time web sessions (e.g., WebRTC meetings, emergency rescue portals) \cite{yue2023low}.

\begin{figure}[t]
    \centering
    \includegraphics[width=\linewidth]{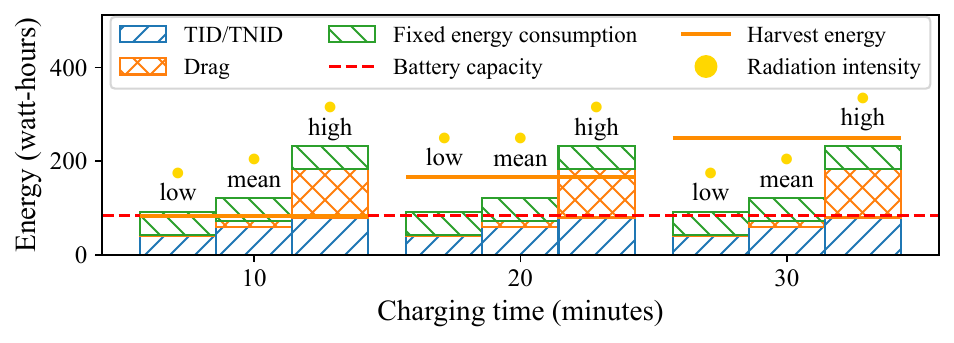} 
    \vspace{-8mm}
    \caption{Energy for radiation recovery cannot be overlooked.}
    \vspace{-5mm}
    \label{fig:solar_intensity_motiv}
\end{figure}

Recovery from these radiation impacts poses dual threats to web services: \emph{interruption} and \emph{unsustainability}. Sleep mode, adopted to mitigate SEE, interrupts active web traffic such as ongoing video conferences, while propulsion to maintain altitude \cite{yu2022simulation} and thermal annealing to repair accumulated damage \cite{srour2013displacement} consume energy that accelerates battery aging. This is a critical issue for LEO satellites with smaller batteries and stronger atmospheric drag. Our energy analysis, using an established LEO energy-harvest/battery model \cite{yang2016towards}, reveals that even low and mean radiation periods consume 50.99\% and 85.87\% of battery capacity for recovery. This leaves insufficient reserves to maintain web services during LEO eclipses (Figure \ref{fig:solar_intensity_motiv}). Inappropriate energy use leads to deep battery discharge (approaching 100\% during eclipses) and extended web downtime, directly eroding user trust in space-based web services. Therefore, adaptively migrating the traffic affected by sleep mode while considering both the energy required for recovery and the impacts on web services is a necessity.

Existing research fails to resolve this tension, as it treats radiation as an afterthought rather than a core design factor. Sustainable routing efforts, which aim to extend battery life for web service stability, focus on solar-aware scheduling or cross-orbit traffic offloading \cite{zeqilai_inorbit_2024} but ignore radiation-induced energy drains that cause unexpected web outages. Robust routing schemes, which prioritize web traffic continuity via redundancy or relays \cite{tao_fast_slow_2023}, cannot prevent radiation-triggered satellite failures that take entire web services offline. Even orbital maneuvers to avoid radiation \cite{zhao2023first} introduce topology instability, raising web request rerouting delays beyond acceptable limits. All these approaches miss the critical link between the space environment and web service resilience.

To fill this gap, we propose a rethink of web service resilience in LEO networks: shifting focus from satellite dynamicity to the space environment’s essential impacts, with radiation as the starting point. We introduce RALT (\textbf{\underline{R}}adiation-\textbf{\underline{A}}ware \textbf{\underline{L}}EO \textbf{\underline{T}}ransmission), a control-plane solution designed explicitly to protect web services from radiation-induced disruptions without hardware modifications. This is a critical advantage for scaling low-cost LEO constellations that support global web access. RALT integrates real-time radiation data (NOAA solar flare indices\footnote{\url{https://ngdc.noaa.gov/stp/solar/solarflares.html}}) and satellite telemetry (battery status, TID/TNID accumulation) to reroute web traffic away from high-risk satellites. Our work shows that unlocking LEO’s potential for global, reliable web service requires rethinking resilience through the lens of the space environment itself.

\begin{figure}[t]
    \mbox{
        \begin{minipage}[t]{0.48\linewidth}
            \centering
            \includegraphics[width=\linewidth]{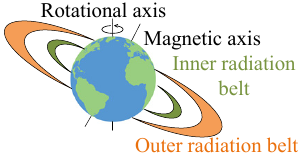} 
        \subcaption{Trapped particles in inner and outer Van Allen Belts.}
    \end{minipage}
    \hspace{2mm}
        \begin{minipage}[t]{0.5\linewidth}
            \centering
           \includegraphics[width=.9\linewidth, height=2.3cm]{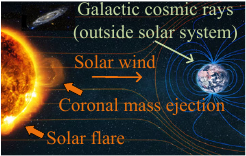} 
            \subcaption{Radiation inside and outside the solar system.} 
        \end{minipage}
    } 
  \vspace{-3.5mm}
      \caption{Classification of typical  radiation sources.} \label{fig:typical_rad}
      \vspace{-5mm}
\end{figure}

\section{Typical Space Radiation Sources} 
As shown in Figure \ref{fig:typical_rad}, the space radiation environment has three dominant source categories \cite{NASASolarEffect}: (1) trapped particles in Earth’s magnetic field (e.g., Van Allen Belts), (2) solar particles from solar flares or coronal mass ejections, and (3) galactic cosmic rays, which are high-energy protons/heavy ions from outside the solar system.

Solar storms directly threaten LEO satellites via transient particle injections and magnetic disturbances, inducing risks like single-event effects and ionization damage. Galactic cosmic rays worsen these risks through four mechanisms \cite{NASASolarEffect}: (1) High-energy penetration (relativistic protons/heavy ions bypass shielding to damage electronics); (2) Cumulative degradation (total ionizing dose effects on semiconductors); (3) Material displacement damage (atomic lattice defects in optics/coatings); (4) Shielding inefficiency (enhanced fluxes during solar minima due to solar-cycle modulation). These threats are amplified by LEO satellites’ limited trapped-particle shielding, as they orbit within the inner Van Allen Belts.

At the network level, radiation-induced satellite failures (e.g., 38 SpaceX satellites’ premature deorbiting during the 2022 geomagnetic storm) and subsequent web service disruptions cause cascading operational impacts. Thus, \emph{solar storms and galactic cosmic rays are the primary radiation threats to space-based web services}.

\section{Fundamentals of RALT}\label{sec:algorithm}
\subsection{Radiation-Aware, Sustainable Rerouting}
\textbf{Basic idea.} At the heart of RALT is a radiation-aware, sustainable rerouting algorithm that first splits LEO satellite network operation into intervals. At each interval’s start, the rerouting mechanism triggers if any satellite on the path of unfinished web service flows is set to enter sleep mode. This occurs either when the satellite is affected by SEE or lacks sufficient energy to mitigate TID/TNID effects and atmospheric drag. The mechanism can be realized through minor adaptations to conventional shortest-path algorithms. For instance, edge weights are redefined as the weighted average of propagation delay (critical for web service responsiveness) and battery life consumption, where the latter is measured via the initial and final Depth-of-Discharge (DoD) of the battery. The final DoD accounts for the energy required to recover from space radiation, and this aligns with cloud providers' focus on scalable, energy-efficient cloud-driven optimization for space networks\footnote{\url{https://aws.amazon.com/ground-station/}}.

We use the following two inequalities to determine when to conduct TID/TNID recovery
\begin{equation}
\label{eq:tid_energy}
\beta_{TID}^{v}(t) \ge \gamma_{TID}\beta_{TID}^{\max},
\end{equation}
\begin{equation}
\label{eq:tnid_energy}
   \beta_{TNID}^{v}(t) \ge \gamma_{TNID}\beta_{TNID}^{\max},
\end{equation}
where $\beta_{TID}^{v}(t)$ and $\beta_{TNID}^{v}(t)$ are the accumulated damage of satellite $v$ at time $t$, while $\beta_{TID}^{\max}$ and $\beta_{TNID}^{\max}$ are the damage thresholds beyond which permanent damage occurs. Two thresholds $\gamma_{TID}\in (0,1)$ and $\gamma_{TNID}\in (0,1)$ are used to determine the time to do annealing for TID/TNID recovery, avoiding abrupt web service interruptions caused by unplanned satellite downtime.

The atmospheric density during radiation events can be obtained from public data (e.g., NOAA solar flare indices) or existing models that estimate density based on radiation intensity \cite{yu2022simulation}. After that, we can obtain the required energy to overcome the radiation-caused atmospheric drag by classical mechanics.

\textbf{Measurement of the TID and TNID accumulated damage.}
While TID impacts are inherently systemic due to the cumulative ionizing energy deposition affecting electronics and power systems, TNID effects, though initially localized to specific components such as optical payloads or RF amplifiers, can propagate to degrade overall mission performance if unmitigated. 

For TID quantification, radiation-hardened sensors are distributed across the satellite to measure cumulative ionizing energy deposition. These measurements are normalized against predefined radiation tolerance limits (e.g., maximum tolerable TID for silicon-based electronics over the satellite’s design lifespan) to compute $\beta_{TID}^{v}(t)$.

TNID quantification, however, requires component-specific displacement damage sensors. These sensors generate a localized TNID value for each monitored component. To obtain the system-level TNID, a weighted aggregation framework is applied. Critical subsystems (e.g., power, communication) are assigned mission-criticality weights $w_i$ and the global TNID value $\beta_{TNID}^v(t)$ is calculated as the weighted sum of the localized values.

\subsection{Control Plane Framework Design}
\begin{figure}[t]
\centerline{\includegraphics[width=0.4\textwidth]{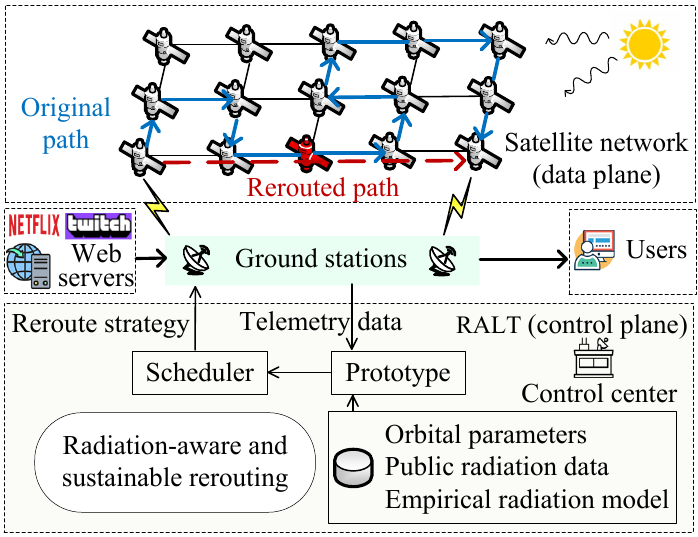}}
\vspace{-2mm}
\caption{Architecture of RALT (control plane).}\label{fig:r3_sat_arc}
\vspace{-4mm}
\end{figure}

The radiation-aware rerouting of RALT operates in the control plane (Figure \ref{fig:r3_sat_arc}), with a core focus on maintaining web service performance and continuity. The control plane integrates a cloud-based control center (hosting a satellite network prototype and scheduler) with distributed ground stations, and operates on the hybrid space-ground infrastructure of cloud web service providers for satellite operations. Satellite networks operate in the data plane.

\textbf{Prototype maintenance.} Initialization leverages two-line element orbital parameters, and ground stations collect telemetry data at predefined intervals to monitor network status, including active web services, battery DoD, and TID/TNID accumulated damage. To enhance the utility of this telemetry data, RALT leverages tools like Amazon EventBridge, which triggers real-time data processing workflows to ensure timely insights into network conditions.

In detail, to derive each satellite's battery DoD, our system leverages legacy telemetry infrastructure through two software stages without hardware modifications. \emph{Stage 1 (DoD embedding)}: The onboard firmware obtains DoD, timestamps the result, and embeds it into standard telemetry packets as backward-compatible auxiliary data, ensuring seamless integration with legacy ground systems. \emph{Stage 2 (Ground validation)}: Ground tools validate telemetry-derived DoD against pre-launch test datasets, applying statistical filters to minimize transmission noise and ensure data reliability.

\textbf{Integration of TID/TNID damage data.} The accumulated damage from TID and TNID effects is derived by extending legacy satellite telemetry systems. More specifically, data from both TID and TNID onboard measurement sensors are transmitted by extending legacy telemetry to ground stations and processed in the \emph{scheduler} of RALT. To integrate TID/TNID data in a backward-compatible way, each sensor periodically computes its metric, timestamps the measured result, and appends it as an auxiliary data field to standard telemetry packets. By preserving the original telemetry structure and simply adding a \emph{radiation} field, existing ground-segment telemetry software can continue decoding packets without modification, enabling runtime adjustments to protect web service flows.

\textbf{Reroute scheduling.} During each interval, the scheduler executes the \emph{radiation-aware and sustainable rerouting} algorithm and, if necessary, generates rerouting strategies. These strategies are then pushed to the satellite networks through ground station control tunnels, providing radiation-resilient transmission services.  

\textbf{Fault tolerance and scalability.} 
Persistent heartbeat monitoring detects ground station failures within 1 minute, initiating telemetry redirection to backup stations. The control center maintains active redundancy across cloud availability zones. New infrastructure (e.g., ground stations or satellites) deployments automatically register through secure handshake protocols, with prototype synchronization achieving consistency within 10 minutes.

\section{Implementation and Evaluation}
\label{sec:eval}
\subsection{Implmentation of RALT}
We implement RALT using about 3,500 lines of Python code and run it on a server equipped with an AMD EPYC
7313 processor. 
  
\textbf{Network properties:} Employing the mapping model of \cite{yu2022simulation} and classical mechanics, we calculate the energy needed to counter increased atmospheric drag. Rerouting paths are scored by a weighted sum of end-to-end latency and battery consumption (weights = 0.5 each). User locations mirror real-world Internet distributions \cite{yang2016towards}, each generating 300 Mb/s of traffic, adjusted for local time. 

\textbf{Radiation-related and energy-related parameters.} We take the radiation intensity and atmospheric density from public sources \cite{RocketandSpaceTechnology}, with sleep-mode probability rising alongside intensity. Two thresholds $\gamma_{TID}$ and $\gamma_{TNID}$ for thermal annealing are both set as $0.7$ \cite{dsouza2021repeated}, with $\beta_{TID}^{\max}=\beta_{TNID}^{\max}=1$. The satellite carries a $5,000$ Wh battery \cite{yang2016towards}; annealing draws 40 W\footnote{\url{https://www.nasa.gov/smallsat-institute/sst-soa/thermal-control/}}, and data transmission uses $0.08$ W·min/Mb \cite{yang2016towards}.

\textbf{Baselines.} We implement the PHOENIX \cite{zeqilai_inorbit_2024} within RALT's scheduler, which also reroutes flows in response to web service interruptions caused by space radiation. However, unlike RALT, it does not account for the energy required for TID/TNID recovery or for overcoming atmospheric drag. We also implement Umbra \cite{tao_fast_slow_2023}. When a satellite in the path of an ongoing flow must be shut down due to space radiation, Umbra adaptively selects a ground station to offload data from the satellite network to the ground.

\subsection{Evaluation Results and Analysis}
\begin{figure}[t]
    \centering
    \includegraphics[width=\linewidth]{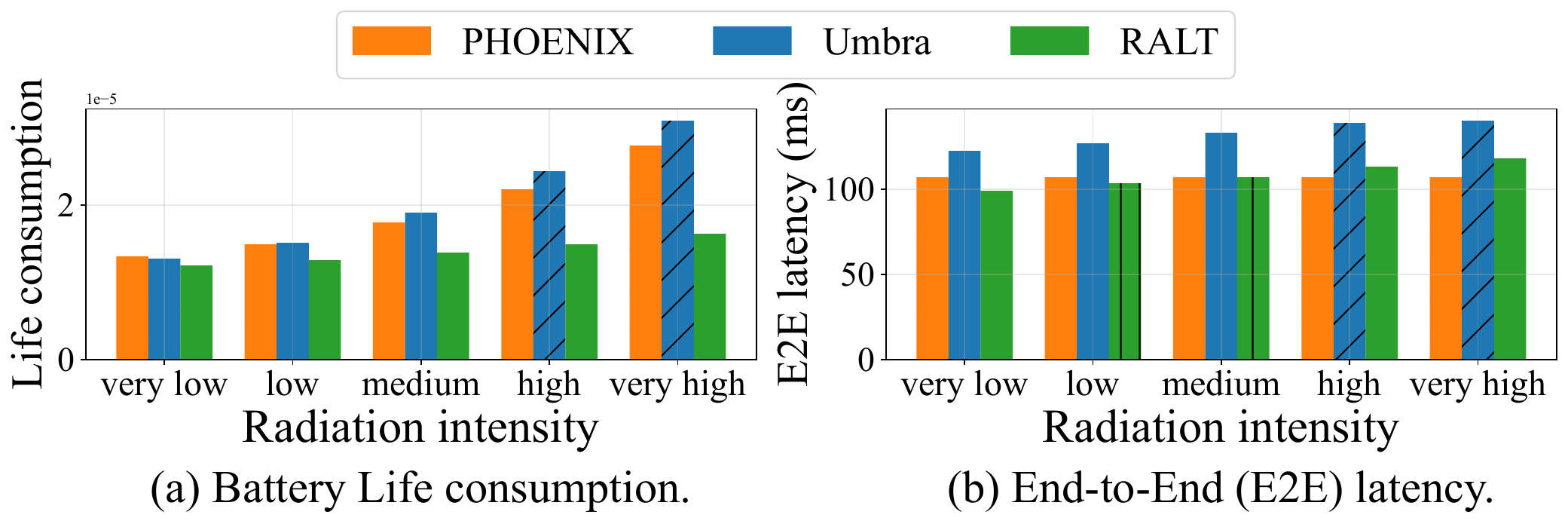} 
    \vspace{-7mm}
    \caption{Performance with different radiation intensities
(Starlink Shell 1: Walker constellation).}
    \vspace{-4mm}
\label{fig:starlink_perf}
\end{figure}

\textbf{Battery life consumption}. Figure \ref{fig:starlink_perf} (a) shows average battery consumption across radiation intensities. RALT consistently outperforms the baselines, with its advantage widening at higher intensities. By incorporating the energy cost of radiation‐induced recovery into its rerouting metric, RALT proactively migrates traffic before deep discharges occur, thereby preserving battery health.

\textbf{Web service End-to-End (E2E) latency.} Figure \ref{fig:starlink_perf} (b) shows that RALT maintains E2E latency comparable to PHOENIX and better than Umbra under all radiation levels. This is achieved by jointly optimizing latency and battery aging during radiation-induced rerouting, smoothing latency fluctuations, and minimizing disruption to web-layer protocols and user-level services (e.g., HTTP/3 requests, dynamic content delivery).

\begin{figure}[t]
    \centering
    \includegraphics[width=\linewidth]{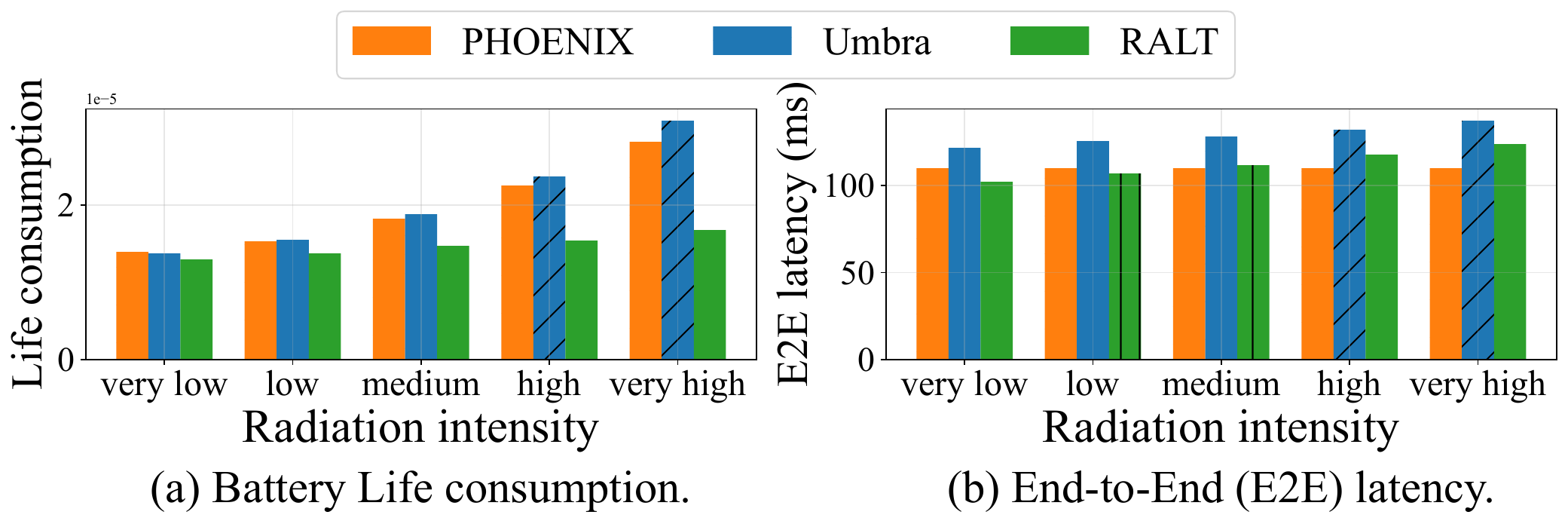} 
    \vspace{-7mm}
    \caption{Performance with different radiation intensities
(Kuiper: near-polar constellation).}
    \vspace{-4mm}
\label{fig:kuiper_perf}
\end{figure}

\begin{figure}[t]
    \centering
    \includegraphics[width=\linewidth]{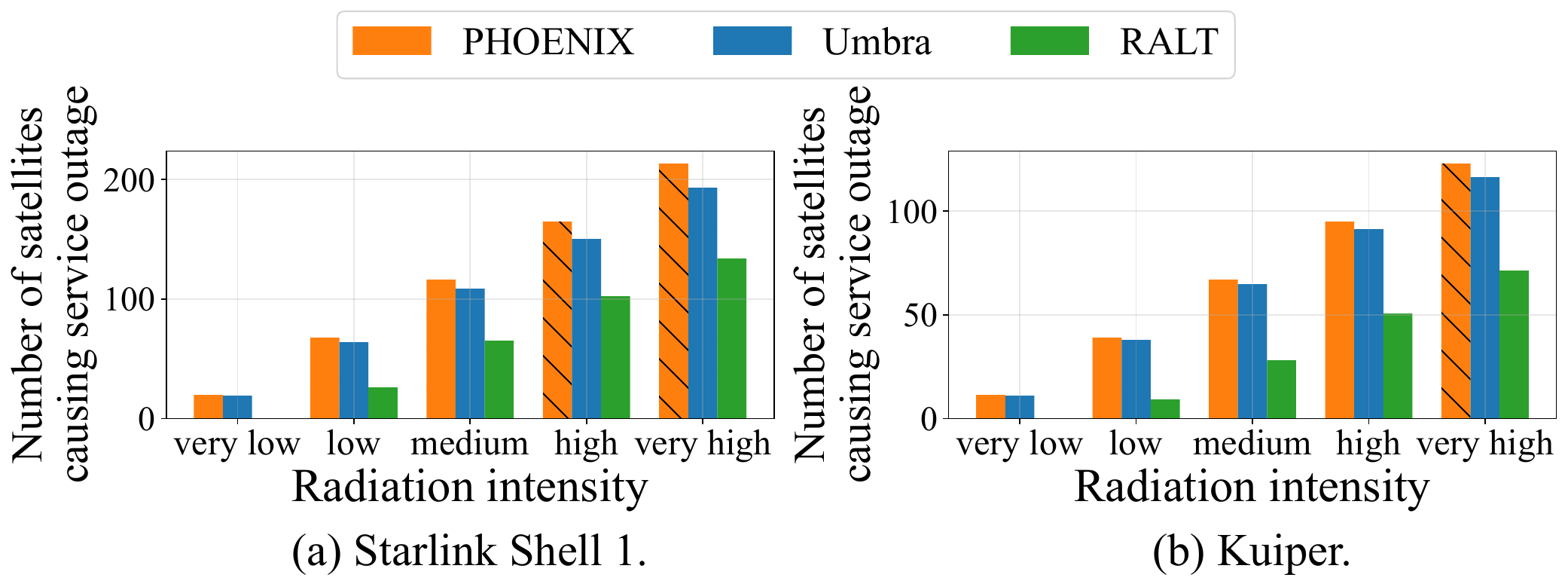} 
    \vspace{-7mm}
    \caption{ Number of satellite shutdowns causing web service
outages/disruptions.}
    \vspace{-4mm}
\label{fig:shut_down_satellite}
\end{figure}

\textbf{Constellation adaptation.} We repeat the evaluation using the Kuiper near-polar constellation. As Figure \ref{fig:kuiper_perf} demonstrates, RALT again reduces battery consumption compared to PHOENIX and Umbra, while sustaining E2E latency similar to that of PHOENIX, underscoring its adaptability to diverse orbital architectures.

\textbf{Number of satellite shutdowns causing web service outages/disruptions.} We count the number of satellites whose depth-of-discharge exceeds 0.95, a threshold that forces them offline to conserve emergency power, and focus on how such shutdowns disrupt web services (e.g., request timeouts, content delivery failures). Experiments across different radiation environments and both satellite constellations show that RALT reduces web service-disrupting satellite shutdowns by 42.95\% on average compared to PHOENIX and Umbra (Figure \ref{fig:shut_down_satellite}). Unlike the baseline approaches, which overlook atmospheric drag and radiation-induced energy costs (TID/TNID recovery), RALT proactively accounts for these factors to prevent deep discharge events that frequently break web service connectivity, increase request failure rates, and accelerate battery degradation. This improvement leads to fewer web service interruptions, more reliable user access and extended battery lifespan, thus enhancing the quality of space-based web services.

\section{Conclusion}
\label{sec:conclusion}
The rapid expansion of cloud-integrated LEO satellite networks is reshaping the landscape. This paper offers a critical rethinking of space-based web services against this backdrop. We introduce RALT, a radiation-aware control-plane solution that addresses the underexplored challenge of network-layer mitigation against radiation risks. By avoiding costly hardware changes and scaling effectively with growing LEO constellations, RALT offers a practical and adaptable approach for next-generation space-based web services. This work advances LEO’s potential to serve as the backbone of ubiquitous, resilient, and low-latency global connectivity.

\section*{Acknowledgments}
We gratefully acknowledge the anonymous reviewers for all of their help and thoughtful comments. H. Wang's work was partly supported by the Scientific Foundation for Youth Scholars of Shenzhen University.

\balance
\bibliographystyle{ACM-Reference-Format}
\bibliography{reference}
\end{document}